\documentstyle[epsfig]{article}

\begin{document}

\title{Vortex Dynamics and the Problem of the Transverse Force in Clean
Superconductors and Fermi Superfluids}
\author{N. B. Kopnin \\
Low Temperature Laboratory, Helsinki University of Technology,\\
P.O. Box 2200, FIN 02015 HUT, Finland\\
and\\
L.D. Landau Institute for Theoretical Physics,\\
Kosygin str. 2, 117940 Moscow, Russia}
\date{}
\maketitle

\begin{abstract}

We review the basic ideas and results on the
vortex dynamics in clean superfluid Fermi systems. The forces acting on 
moving vortices are discussed including the problem of the transverse force
which was a matter of confusion for quite a time. Finally, we formulate the 
equations of the vortex dynamics which include all the forces and the inertial 
term associated with excitations bound to the moving vortex. 

\end{abstract}

\section{Introduction}

\label{sect/Boleq}

Study of the vortex dynamics opens promising perspectives for understanding the
most fundamental properties of condensed matter especially 
for superconductors and superfluids. Among them,
clean systems, i.e., the systems where the mean free path of quasiparticles
is much longer than the characteristic coherence length, offer more intriguing 
physics than dirty superconductors. For
example, one of the fundamental problems can be formulated as follows:
Speaking of clean superconductors one can, in particular, think of such a
system where no relaxation processes are available, i.e., the mean free path
of excitations is infinite. In particular, vortices in such systems 
should move without
dissipation since there is no mechanism to absorb the energy. The vortex
velocity should then be parallel to the transport current, which makes the
induced electric field perpendicular to the current; the dissipation thus 
vanishes $%
{\bf j}\cdot {\bf E}=0$. This contrasts with what we know of dirty
superconductors: the vortex motion there is dissipative, and each moving
vortex experiences a large friction; it generates an electric field parallel
to the transport current which produces energy dissipation. Clearly, a
crossover should occur from dissipative to non--dissipative vortex motion as
the quasiparticle mean free path increases. The question is what is the
condition which controls the crossover?

This question is of a fundamental importance for our understanding of the
dynamics of superconductors and, in a more broad sense, for the
understanding of dynamic properties of quantum condensed matter in general.
To illustrate the problem one can consider a simple example as follows. One
can argue that a time dependent non--dissipative superconducting state,
similarly to any other quantum state, can be described by a Hamiltonian
dynamics based on a time dependent Schr{\"{o}}dinger equation%
\index{Schroedinger equation}. Such a description for a weakly interacting
Bose gas has been suggested by Pitaevskii and Gross \cite{Pitaevskii,Gross}. 
It is widely used for superfluid helium II as well. The
Gross-Pitaevskii equation%
\index{Gross--Pitaevskii equation} is essentially a nonlinear
Schr\"{o}dinger equation, it has the imaginary factor $-i\hbar $ in front of
the time derivative of the condensate%
\index{Bose!condensate} wave function $\partial \Psi /\partial t$. On the
other hand, the time dependent Ginzburg--Landau model%
\index{Ginzburg--Landau!time dependent model} which is a particular case of
a more general Model $F$ dynamics \cite{Halperin} is believed to describe a
relaxation dynamics of superconductors near the transition temperature. In
contrast to the Gross--Pitaevskii equation, it has the time derivative $%
\partial \Psi /\partial t$ with a real factor in front of it. The question
which we are interested in can be formulated as follows: What is the
condition when the imaginary prefactor transforms into a real one?

It seems that there is no universal answer to this simple question in
general. However, the problem of crossover from non--dissipative to
dissipative behavior of a condensed matter state can be solved for the
particular example of vortex dynamics. It is known that the relaxation
constant in the time dependent Ginzburg--Landau model has in fact a small
imaginary part \cite{Fukuyama,Dorsey,KIK} which results in appearance of a
small transverse component of the electric field with respect to the
current. We shall see later that the transverse component of the electric
field increases at the expense of the longitudinal component as the mean
free path of excitations grows. The crossover condition, however, does not
coincide simply with the condition which divides superconductors between
dirty and clean ones. The criterion rather involves the spectrum%
\index{energy spectrum} of excitations localized in the vortex cores%
\index{excitations!in the vortex core}; the distance between their levels
takes the part of the energy gap. The condition for a non--dissipative
vortex motion requires that the relaxation%
\index{relaxation} rate of localized excitations is smaller than the
distance between the levels. This implies a much longer mean free path%
\index{mean free path} of excitations than the condition for a
superconductor to be just in a clean limit.

In the present paper we review the basic ideas and results on the
vortex dynamics in clean Fermi systems. We concentrate largely on 
superconductors,
the case of superfluid $^3$He can be easily incorporated if one takes the 
limit of zero charge of carriers in the final results (to be discussed it in 
more detail later). We consider forces acting on moving vortices and
clarify the conditions for the crossover from dissipative to non--dissipative
vortex dynamics. A great deal of attention is given to the problem of the
so called transverse force, i.e., the force perpendicular to the vortex 
velocity, which was a matter of confusion for quite a time since its
discovery. Finally, we formulate the equations of the vortex dynamics
which include all the forces and the inertial term associated with excitations
bound to the moving vortex. 

\section{Boltzmann kinetic equation approach}

The forces on a vortex come from several different sources
including the hydrodynamic Magnus force, the force produced by excitations
scattered from the vortex, and the force associated with the momentum flow
from the heat bath to the vortex through the localized excitations. The
whole rich and exciting physics involved in the vortex dynamics can be
successfully described by the general microscopic time dependent theory.
Here we rather discuss a general picture using a simple semi--classical
approach based on the Boltzmann kinetic equation%
\index{kinetic equation!Boltzmann}. The semi--classical approach assumes that
the wavelength of excitations is much shorter than the superfluid coherence
length, $p_F\xi \gg 1$ (from now on we put $\hbar =1$). This condition 
is quite safely fulfilled in almost
all superconductors and in superfluid $^3$He. Only in some high--temperature
superconductors, its accuracy may be not very good. For simplicity, we
consider $s$--wave superconductors%
\index{superconductors!s-wave}. We concentrate on isolated vortices such
that their cores do not overlap, i.e., on the region of magnetic fields $%
H\ll H_{c2}$.

One can distinguish two types of excitations: Excitations localized in the
vortex core, and those which are not localized but move in the vortex
potential under the action of magnetic filed. We start our discussion with
the localized excitations.

\subsection{Localized excitations}

\begin{figure}[t]
\centerline{\epsfxsize= 6.5cm\epsfbox{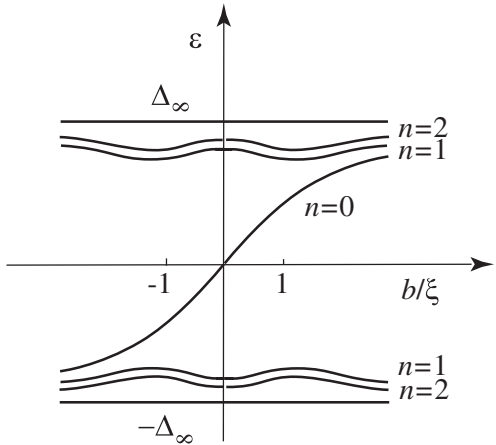}}
\caption{Spectrum of excitations localized in the vortex core.
}
\label{fig-levels}
\end{figure}

We remind that the profile of the order parameter $\Delta \left( {\bf r}%
\right) $ near the vortex core%
\index{vortex core} produces a potential well where localized states%
\index{vortex core!quasiparticle states in} with a discrete spectrum exist.
The localized states correspond to energies $|\epsilon |<\Delta _{\infty }$.
The spectrum has the so called anomalous branch with the radial quantum
number $n=0$ whose energy varies from $-\Delta _{\infty }$ to $+\Delta
_{\infty }$ as the particle impact parameter $b$ changes from $-\infty $ to $%
+\infty $ and crosses $\epsilon =0$, being an odd function of $b$. For low $%
\epsilon \ll \Delta _{\infty }$, the anomalous branch is $\epsilon
_{0}=-\omega _{0}\mu $ where $\mu =-bp_{\perp }$ is the angular momentum,
and $p_{\perp }$ is the momentum in the plane perpendicular to the vortex
axis \cite{Caroli} (see also Fig. \ref{fig-levels}). In a $s$--wave
superconductor with an axisymmetric vortex, the angular momentum $\mu $ is
quantized and so is the spectrum, $\omega _{0}$ being the distance between
the discrete levels in the vortex core. The spectrum also has branches with $%
n\ne 0$ which are separated from the one with $n=0$ by energies of the order
of $\Delta $. They are even in $b$ and symmetric in energy with respect to $%
\epsilon =0$. We denote the separation between the levels with neighboring
angular momenta through 
\begin{equation}
\omega _{n}=p_{\perp }^{-1}%
\frac{\partial \epsilon _{n}}{\partial b}=-\frac{\partial \epsilon _{n}}{%
\partial \mu }.  \label{omegar}
\end{equation}
The interlevel spacing $\omega _{n}(b)$ is an even function of $b$ for $n=0$%
, and it is an odd function for $n\ne 0$.

\begin{figure}[t]
\centerline{\epsfxsize= 8.5cm\epsfbox{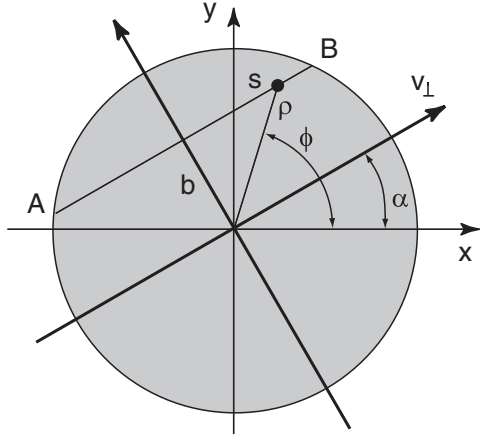}}
\caption{Transformation from cylindrical coordinates to the frame
associated with a particle moving in the vortex core. The line AB is the 
particle trajectory making an angle $\alpha$ with the $x$ axis and
passing through the position point $(\rho ,\phi )$
at a distance $b$ from the vortex axis.
}
\label{vort-coor}
\end{figure}

We choose the direction of the $z$ axis in such a way that the vortex has a
positive circulation. The $z$ axis is thus parallel to the magnetic field for
positive charge of carriers, and it is antiparallel to it for negative
charge: $\hat{{\bf z}}=\hat{{\bf h}}\,{\rm sign}\,(e)$. Since the particle
velocity ${\bf v}_{\perp }$ in the plane perpendicular to the vortex axis
makes an angle $\alpha $ with the $x$ axis, the
cylindrical coordinates of the position point $(\rho ,\phi )$ are connected
with the impact parameter and the coordinate along the trajectory through $%
\rho ^{2}=b^{2}+s^{2}$ where 
\begin{equation}
b=\rho \sin (\phi -\alpha )~;~s=\rho \cos (\phi -\alpha ).  \label{b-phi}
\end{equation}
The coordinates are shown in Fig. \ref{vort-coor}.

The first step is as follows. We assume that the quasiclassical spectrum $%
\epsilon _{n}\left( b\right) $ of a particle plays the role of its effective
Hamiltonian%
\index{Hamiltonian}. We can thus invoke the Boltzmann equation in the
canonical form 
\begin{equation}
\frac{\partial {\sl f}}{\partial t}+\frac{\partial {\sl f}}{\partial \alpha }%
\, \frac{\partial \epsilon _{n}}{\partial \mu }-\frac{\partial \epsilon
_{n}}{\partial \alpha }\, \frac{\partial {\sl f}}{\partial \mu }=\left( 
\frac{\partial {\sl f}}{\partial t}\right) _{{\rm coll}},
\label{Boltzmann/eq1}
\end{equation}
to describe the quasiparticle distribution \cite{Stone}. Equation (\ref
{Boltzmann/eq1}) has been derived in Ref. \cite{BlatterKop} from the set of
microscopic kinetic equations.

In the time derivative, the energy $\epsilon _{n}$ contains a time
dependence through $\mu (t)=\left[ ({\bf r}-{\bf v}_{L}t)\times {\bf p}%
\right] \cdot \hat{{\bf z}}$ such that 
\begin{equation}
\frac{\partial {\sl f}^{\left( 0\right) }}{\partial t}=\frac{\partial {\sl f}%
^{(0)}}{\partial \epsilon }\frac{\partial \epsilon _{n}}{\partial \mu }\frac{%
\partial \mu }{\partial t}=\frac{\partial {\sl f}^{(0)}}{\partial \epsilon }%
\frac{\partial \epsilon _{n}}{\partial \mu }\left( [{\bf p}_{\perp }\times 
{\bf v}_{L}]\cdot \hat{{\bf z}}\right) .
\end{equation}
Kinetic equation takes the form 
\begin{equation}
\frac{\partial {\sl f}_1}{\partial t}+ \frac{\partial {\sl f}^{(0)}}{%
\partial \epsilon }\frac{\partial \epsilon _{n}}{\partial \mu }\left( [{\bf p%
}_{\perp }\times {\bf v}_{L}]\cdot \hat{{\bf z}}\right)+ \frac{\partial {\sl %
f}_1}{\partial \alpha }\frac{\partial \epsilon _{n}}{\partial \mu }-\frac{%
\partial \epsilon _{n}}{\partial \alpha }\frac{\partial {\sl f}_1}{\partial
\mu }=\left( \frac{\partial {\sl f}}{\partial t}\right) _{{\rm coll}},
\label{kin/local}
\end{equation}
where the distribution function is separated into an equilibrium and
a non--equilibrium parts, ${\sl f}={\sl f}^{(0)}+{\sl f}_1$, respectively.
Here ${\sl f}^{(0)}=1-2n_\epsilon =\tanh \left( \epsilon /2T\right)$
with $n_\epsilon$ being the Fermi function.

We shall simplify the collision integral in Eq. (\ref
{Boltzmann/eq1}) using the relaxation--time approximation%
\index{collision integral!tau--approximation for} 
\begin{equation}
\left( 
\frac{\partial {\sl f}}{\partial t}\right) _{{\rm coll}}=-\frac{{\sl f}_{1}}{%
\tau _{n}}  \label{tauapprox}
\end{equation}
where $\tau _{n}\sim \tau $. With this approximation, the mean free time
can be of any origin. For definiteness, we assume that the most effective
relaxation is brought about by impurities, as is the case in almost
all practical superconducting compounds.  However, the 
relaxation time $\tau$ can be due to the electron--phonon interaction, 
as well, 
or due to quasiparticle--quasiparticle scattering. The latter is, in fact, 
the only relaxation mechanism available in $^3$He \cite{KL2tau}.
Therefore, the term ``clean'' does not necessarily mean a low concentration
of impurities but simply refers to a situation when the mean free time is long.

Equation (\ref{Boltzmann/eq1}) with the collision integral in the form of
Eq. (\ref{tauapprox}) is easy to solve. For an axi--symmetric $s$%
-wave vortex the energies $\epsilon _{n}$ do not depend on $\alpha $ and the
term $\partial \epsilon _{n}/\partial \alpha $ vanishes. Let us take the
distribution function in the form 
\begin{equation}
{\sl f}_{1}=-{\frac{{\partial {\sl f}^{(0)}}}{{\partial \epsilon }}}\left[ {%
\gamma _{{\rm O}}}([{\bf v}_{L}\times {\bf p}_{\perp }]\hat{\cdot {\bf z}})+{%
\gamma _{{\rm H}}}({\bf v}_{L}\cdot \,{\bf p}_{\perp })\right]
\label{f1loc/sph}
\end{equation}
where the factors $\gamma _{{\rm O,H}}$ are to be found from the Boltzmann
equation (\ref{Boltzmann/eq1}). The result for a steady vortex motion is 
\begin{equation}
\gamma _{{\rm O}}=\frac{\omega _{n}\tau _{n}}{\omega _{n}^{2}\tau _{n}^{2}+1}%
,\qquad \gamma _{{\rm H}}=\frac{\omega _{n}^{2}\tau _{n}^{2}}{\omega
_{n}^{2}\tau _{n}^{2}+1}.  \label{gammas}
\end{equation}
This generalizes the result first obtained in Ref. \cite{KopKr1}.

\subsection{Delocalized excitations}

A delocalized particle moves mostly far from the vortex core where the order
parameter is constant and the superfluid velocity potential ${\bf pv}_{s}$
is small compared to $\Delta $. Kinetic equation for delocalized excitations
can thus be written as for a particle in a magnetic field with a
semi--classical spectrum 
\begin{equation}
\epsilon _{p}=\sqrt{\xi _{p}^{2}+\Delta _{\infty }^{2}}  \label{qcspectrum}
\end{equation}
where $\xi _{p}=p^{2}/2m-E_{F}$. As shown in Ref. \cite{KV/magspectr}
the kinetic equation for a particle moving in a vortex array
has the conventional form  
\[
\frac{\partial {\sl f}}{\partial t}+\frac{\partial {\sl f}}{\partial {\bf p}}%
\cdot {\bf f}+{\bf v}_{g}\cdot \frac{\partial {\sl f}}{\partial {\bf r}}%
=\left( \frac{\partial {\sl f}}{\partial t}\right) _{{\rm coll}} .
\]
The force is the elementary Lorentz force 
\begin{equation}
{\bf f}=\frac{\partial {\bf p}}{\partial t}=\frac{e}{c}{\bf v}_{g}\times 
{\bf H=}\frac{\omega _{c}}{g}\left[ {\bf p}_{F}\times {\bf \hat{z}}\right]
\label{dpdt-deloc}
\end{equation}
where $\omega _{c}=\left| e\right| H/mc$ is the cyclotron frequency, 
\begin{equation}
{\bf v}_{g}=\frac{\partial \epsilon _{p}}{\partial {\bf p}}=\frac{{\bf v}_{F}%
}{g}  \label{v-group}
\end{equation}
is the group velocity, and 
\[
g=\frac{\epsilon }{\sqrt{\epsilon ^{2}-\Delta _{\infty }^{2}}} .
\]
The driving term can be written as 
\[
\frac{\partial {\sl f}}{\partial t}=\frac{\partial {\sl f}^{\left( 0\right) }%
}{\partial \epsilon }\frac{\partial {\sl \epsilon }}{\partial t} 
\]
where 
\[
\frac{\partial {\sl \epsilon }}{\partial t}=e{\bf v}_{g}\cdot {\bf E=}\frac{e%
}{c}{\bf v}_{g}\cdot \left[ {\bf H}\times {\bf v}_{L}\right] =\frac{\omega
_{c}}{g}{\bf p}_{\perp }\cdot \left[ {\bf \hat{z}}\times {\bf v}_{L}\right] .
\]
The kinetic equation becomes 
\begin{equation}
\frac{\partial {\sl f}_{1}}{\partial t}+\frac{\partial {\sl f}^{\left(
0\right) }}{\partial \epsilon }\frac{\omega _{c}}{g}{\bf p}_{\perp }\cdot
\left[ {\bf \hat{z}}\times {\bf v}_{L}\right] +\frac{\partial {\sl f}_{1}}{%
\partial {\bf p}}\cdot \frac{\omega _{c}}{g}\left[ {\bf p}_{F}\times {\bf 
\hat{z}}\right] =\left( \frac{\partial {\sl f}}{\partial t}\right) _{{\rm %
coll}} . \label{kin/deloc}
\end{equation}
We omit the spatial derivative of the distribution function since it is
constant in space.

For the energy spectrum of Eq. (\ref{qcspectrum}) the collision integral 
is \cite{AG} 
\[
\left( \frac{\partial {\sl f}}{\partial t}\right) _{{\rm coll}}=-\frac{1}{%
g\tau }{\sl f}_{1} .
\]
Kinetic equation (\ref{kin/deloc}) takes the final form 
\[
{\bf p}_{\perp }\cdot \left[ {\bf \hat{z}}\times {\bf v}_{L}\right] \frac{%
\partial {\sl f}^{\left( 0\right) }}{\partial \epsilon }-\frac{\partial {\sl %
f}_{1}}{\partial \alpha }=-\frac{1}{\omega _{c}\tau }{\sl f}_{1} .
\]
Its solution is Eq. (\ref{f1loc/sph}) with 
\begin{equation}
\gamma _{{\rm O}}^{\prime }=\frac{\omega _{c}\tau }{\omega _{c}^{2}\tau
^{2}+1},\qquad \gamma _{{\rm H}}^{\prime }=\frac{\omega _{c}^{2}\tau ^{2}}{%
\omega _{c}^{2}\tau ^{2}+1}. \label{gammas/del}
\end{equation}

\section{Forces}

At the second step we calculate the force acting on a vortex from the
environment. This force is exerted via excitations which travel near the
vortex through transfer of their momenta to the vortex. Consider first
localized excitations. The transferred momentum is 
\begin{equation}
{\bf F}_{{\rm env}}^{\left( {\rm loc}\right) }=\frac{1}{2}\sum_{n}\int \frac{%
dp_{z}}{2\pi }\,\frac{d\alpha d\mu }{2\pi }\,\frac{\partial {\bf p}_{n}}{%
\partial t}\,{\sl f}_{1}=-\frac{1}{2}\sum_{n}\int \frac{dp_{z}}{2\pi }\,%
\frac{d\alpha d\mu }{2\pi }\,\frac{\partial \epsilon _{n}}{\partial b}[\hat{%
{\bf z}}\times \hat{{\bf v}}_{\perp }]\,{\sl f}_{1}.  \label{force/boltz}
\end{equation}
Here we make use of the Hamilton equation 
\begin{equation}
\frac{\partial {\bf p}_{n}}{\partial t}=-\nabla \epsilon _{n}=-\frac{%
\partial \epsilon _{n}}{\partial b}[\hat{{\bf z}}\times \hat{{\bf v}}_{\perp
}].
\end{equation}
The second equality follows from the fact that, in the coordinate frame $%
\left( s,b\right) $ of Fig. \ref{vort-coor}, the energy only depends on the 
particle impact parameter $b$. 

The transfer of the momentum from delocalized excitations in the
semi--classical approximation has the form 
\[
{\bf F}_{{\rm env}}^{\left( {\rm del}\right) }=\frac{1}{2}\sum_{n}\int \frac{%
dp_{z}}{2\pi }\,\int \frac{d\alpha }{2\pi }\,\frac{\partial {\bf p}_{n}}{%
\partial t}\,{\sl f}_{1} 
\]
where $\partial {\bf p_{n}}\partial t$ is given by Eq. (\ref{dpdt-deloc}).
The force becomes 
\[
{\bf F}_{{\rm env}}^{\left( {\rm del}\right) }=\frac{1}{2}\sum_{n}\int \frac{%
dp_{z}}{2\pi }\,\int \frac{d\alpha }{2\pi }\,\left[ {\bf p}_{F}\times {\bf 
\hat{z}}\right] \frac{\omega _{c}}{g}\,{\sl f}_{1}. 
\]
The sum is over quantized states of a semi--classical quasiparticle in a
magnetic field. Indeed, the energy spectrum Eq. (\ref{qcspectrum}) becomes
quantized \cite{KV/magspectr} in a magnetic field and turns into a modified
Landau spectrum $\epsilon _{n}$ such that 
\[
\sqrt{\epsilon _{n}^{2}-\Delta _\infty ^{2}}=\omega _{c}n-
{\bf p}_{\perp }^{2}/2m. 
\]
The quantity 
\[
\frac{g}{\omega _{c}}=\frac{\partial n}{\partial \epsilon } 
\]
is thus the density of states. As a result, 
\[
\sum_{n}\rightarrow \int \frac{\partial n}{\partial \epsilon }\,d\epsilon 
\]
and the force becomes \cite{KL} 
\begin{equation}
{\bf F}_{{\rm env}}^{\left( {\rm del}\right) }=\int_{\epsilon >\Delta_\infty}
d\epsilon \int \frac{dp_{z}}{2\pi }\,\int \frac{d\alpha }{2\pi }\,
\left[ {\bf p}_{\perp }\times {\bf \hat{z}}\right] \,{\sl f}_{1} .
\label{force/boltz/del}
\end{equation}

The total force is 
\begin{equation}
{\bf F}_{{\rm env}}={\bf F}_{{\rm env}}^{\left( {\rm loc}\right) }+{\bf F}_{%
{\rm env}}^{\left( {\rm del}\right) } . \label{force/total}
\end{equation}
With the Ansatz (\ref{f1loc/sph}), the total force Eq. (\ref{force/total})
splits into two terms ${\bf F}_{{\rm env}}={\bf F}_{\parallel }+{\bf F}%
_{\perp }$, with the friction%
\index{friction force} ${\bf F}_{\parallel }$ and transverse%
\index{transverse force} ${\bf F}_{\perp }$ forces given by 
\begin{eqnarray}
{\bf F}_{\parallel } &=&-\pi N\left[ \left\langle \!\!\!\left\langle
\sum_{n}\int \omega _{n}\gamma _{{\rm O}}\,%
\frac{\partial {\sl f}^{(0)}}{\partial \epsilon }\,\frac{d\mu }{2}%
\right\rangle \!\!\!\right\rangle +\left( 1-\tanh \frac{\Delta _{\infty }}{2T%
}\right) \gamma _{{\rm O}}^{\prime }\right] {\bf v}_{L},  \label{force/par}
\\
{\bf F}_{\perp } &=&\pi N\left[ \left\langle \!\!\!\left\langle \int \omega
_{0}\gamma _{{\rm H}}\,\frac{\partial {\sl f}^{(0)}}{\partial \epsilon }\,%
\frac{d\mu }{2}\right\rangle \!\!\!\right\rangle +\left( 1-\tanh \frac{%
\Delta _{\infty }}{2T}\right) \gamma _{{\rm H}}^{\prime }\right] \left[ {\bf 
\hat{z}\times v}_{L}\right] ,  \label{force/tr}
\end{eqnarray}
where $N$ is the quasiparticle (electron) density%
\index{electron!density}, $\left\langle \!\left\langle \ldots \right\rangle
\!\right\rangle $ is the average over the Fermi surface with the weight $\pi
p_{\perp }^{2}$, 
\begin{equation}
\left\langle \!\left\langle \ldots \right\rangle \!\right\rangle
=V_{F}^{-1}\int \pi p_{\perp }^{2}dp_{z}\left( \ldots \right) ,
\label{aver/FS}
\end{equation}
and $V_{F}$ is the volume encompassed by the Fermi surface. For an isotropic
Fermi surface, 
\[
\left\langle \!\left\langle \ldots \right\rangle \!\right\rangle =%
\frac{3}{4}\int \sin ^{3}\theta \,d\theta \,\left( \ldots \right) . 
\]
where $p_{z}=p_{F}\cos \theta $. Only the spectral branch with $n=0$
contributes to the transverse force ${\bf F}_{\perp }$ as all $\omega _{n}$
with $n\ne 0$ are odd functions of $\mu $ and thus drop out of the sum over $%
n$ in Eq. (\ref{force/tr}).

We emphasize that the force ${\bf F}_{{\rm env}}$ is defined as the response
of the whole environment to the vortex displacement. It is therefore the 
{\em total} force acting on the vortex from the ambient system, including
all partial forces such as the longitudinal friction force and the
non-dissipative transverse force. The transverse force, in turn, includes
various parts which can be identified historically \cite{Sonin} as the
Iordanskii force \cite{Iordanskii}, the spectral flow force \cite
{KopVol/spflow,specflow,specflow1}. and the Magnus force. We
shall discuss this later in more detail in Section \ref{sect/forces-disscus}.

\subsection{Flux flow conductivity}

The force from the environment is balanced by the
Lorentz force%
\index{Lorentz force}: 
\begin{equation}
{\bf F}_{{\rm L}}={%
\frac{{\Phi _{0}}}{c}}[{\bf j}_{{\rm tr}}\times \hat{{\bf z}}]{\rm sign}%
\,(e),  \label{Lorentzforce1}
\end{equation}
with the flux quantum $\Phi _{0}=\pi c/|e|$. The force balance equation 
\begin{equation}
{\bf F}_{{\rm L}}+{\bf F}_{{\rm env}}=0 \label{forcebalance}
\end{equation}
determines the transport current in terms of the vortex velocity and thus
allows us to find the flux flow conductivity tensor. The {\it longitudinal
force} ${\bf F}_{\parallel }$ defines the friction coefficient in the vortex
equation of motion and determines the Ohmic component of the conductivity%
\index{conductivity!Ohmic} $\sigma _{{\rm O}}$. Expressing the vortex
velocity ${\bf v}_{L}$ through the average electric field ${\bf E}$, as $%
{\bf v}_{L}=c\,[{\bf E}\times 
\hat{{\bf z}}]/B\,{\rm sign}(e)$, we find 
\begin{equation}
\sigma _{{\rm O}}=\frac{N|e|c}{B}\left[ \left\langle \!\!\!\left\langle
\sum_{n}\int \frac{\omega _{n}\tau _{n}}{\omega _{n}^{2}\tau _{n}^{2}+1}\,%
\frac{\partial {\sl f}^{(0)}}{\partial \epsilon }\,\frac{d\epsilon }{2}%
\right\rangle \!\!\!\right\rangle +\left( 1-\tanh \frac{\Delta _{\infty }}{2T%
}\right) \frac{\omega _{c}\tau }{\omega _{c}^{2}\tau ^{2}+1}\right] .
\label{sigma-O}
\end{equation}
The {\it transverse force} determines the Hall conductivity%
\index{conductivity!Hall} 
\begin{equation}
\sigma _{{\rm H}}=%
\frac{Nec}{B}\left[ \left\langle \!\!\!\left\langle \int \frac{\omega
_{0}^{2}\tau _{0}^{2}}{\omega _{0}^{2}\tau _{0}^{2}+1}\,\frac{\partial {\sl f%
}^{(0)}}{\partial \epsilon }\,\frac{d\epsilon }{2}\right\rangle
\!\!\!\right\rangle +\left( 1-\tanh \frac{\Delta _{\infty }}{2T}\right) 
\frac{\omega _{c}^{2}\tau ^{2}}{\omega _{c}^{2}\tau ^{2}+1}\right] .
\label{sigma-H}
\end{equation}

The main conclusion is that the Ohmic and Hall conductivities depend on the
purity of the sample through the parameters $\omega _{0}\tau $ (remind that $%
\omega _{n}\sim \omega _{0}$) and $\omega _{c}\tau $. Note that $\omega
_{c} \sim (H/H_{c2})\omega _{0} $ thus $\omega _c \ll \omega _0$.
One can distinguish two regimes: moderately clean $\omega _{0}\tau \ll 1$
and superclean $\omega _{0}\tau \gg 1$. Note that the moderately clean
regime still requires that the superconductor is clean in the usual sense $%
\Delta _{\infty }\tau \gg 1$.

In the moderately clean limit where $\omega _{0}\tau \ll 1$, the
conductivity roughly follows the Bardeen and Stephen expression
\cite{BS} at low temperatures though it exhibits an
extra temperature-dependent factor $\Delta _{\infty }/T_{c}$ on approaching $%
T_{c}$ \cite{KL} 
\[
\sigma _{{\rm O}}\sim \sigma _{n}%
\frac{H_{c2}}{H}\frac{\Delta _{\infty }}{T_{c}}. 
\]
The factor $\Delta _{\infty }/T_{c}$ appears because the number of
delocalized quasiparticles contributing to the vortex dynamics decreases
near $T_{c}$.  This extra factor has been recently identified experimentally
\cite{Kambe}. The Hall conductivity and the Hall angle
are small $\sigma _{{\rm H}}\sim \left( \omega _{0}\tau
\right) \sigma _{{\rm O}}$ and $\tan \Theta _{H}\sim \left( \omega _{0}\tau
\right) $, respectively. The contribution from delocalized states are not
important since $\omega _{c}\ll \omega _{0}$.

In the superclean limit $\omega _0\tau \gg 1$, on the contrary, the Ohmic
conductivity is small. The vortex dynamics becomes non--dissipative. In
particular, if $\omega _c\tau \ll 1$, the corresponding Hall conductivity%
\index{conductivity!Hall} is 
\[
\sigma _{H}=%
\frac{Nec}{B}\tanh \left( \frac{\Delta _{\infty }}{2T}\right) . 
\]
If $\omega _c\tau \gg 1$, the hyperbolic tangent should be replaced with
unity.

\section{Transverse force\label{sect/forces-disscus}}

Let us now discuss the forces acting on a moving vortex in more detail. The
friction force%
\index{friction force} is determined by the Eq. (\ref{force/par}). It is
proportional to the mean free path of excitations in the moderately clean
regime and vanishes in the superclean limit. The transverse force%
\index{transverse force} Eq. (\ref{force/tr}) deserves a more careful
discussion because it has been a matter of controversy for a long time since
first calculated for vortices in helium II by Lifshitz and Pitaevskii \cite
{LifPit} and then by Iordanskii \cite{Iordanskii}.  The review
\cite{Sonin} tells about old disputes (see also \cite{Sonin-Iord}). Recently, 
the presence of the transverse
force has been questioned in Refs. \cite{TAN,Ao}.

As we see, however, the microscopic picture gives a finite transverse force
in a full accordance with the symmetry arguments which allow a transverse
force in a chiral system such as a moving vortex in presence of a superflow.
Of course, the magnitude of the transverse force is not an universal 
quantity. It appears to depend on parameters of the system. Indeed,
in the superclean limit, when both $\omega _{c}\tau $ and $\omega _{0}\tau $
are much larger than unity, the factors $\gamma _{{\rm H}}=\gamma _{{\rm H}%
}^{\prime }=1$, and the transverse force Eq. (\ref{force/tr}) becomes 
\begin{equation}
{\bf F}_{\perp }={\pi }N[%
\hat{{\bf z}}\times {\bf v}_{L}].  \label{Ftransv/superclean}
\end{equation}
The balance Eq. (\ref{forcebalance}) gives the transport current in the form 
\begin{equation}
{\bf j}_{{\rm tr}}=Ne{\bf v}_{L}.\label{jtr/superclean}
\end{equation}
This equation is consistent with the Helmholtz theorem of conservation of
circulation in an ideal fluid: vortices move together with the flow. The
correction to the distribution function of excitations is simply 
\begin{equation}
{\sl f}_{1}=-{\frac{{\partial {\sl f}^{(0)}}}{{\partial \epsilon }}}({\bf v}%
_{L}\cdot \,{\bf p}_{\perp }) \label{distr/ideal}
\end{equation}
such that the full distribution function is 
${\sl f}={\sl f}^{\left( 0\right) }\left( \epsilon -{\bf v}%
_{L}\cdot \,{\bf p}\right) $: excitations move together with the vortex being
in equilibrium with the vortex
array. To understand this let us consider first the mean free path of
delocalized excitations with respect to their collisions with vortices. If
the vortex cross section is $\sigma _{v}$, the mean free path is 
$ \ell _{v}=1/\sigma _{v}n_{v}$ 
where $n_{v}=B/\Phi _{0}$ is the density of vortices. We shall see in a
moment that the vortex cross section is $\sigma _{v}\sim p_{F}^{-1}$ so that 
\begin{equation}
\ell _{v}\sim p_F/n_v \sim v_{F}/\omega _{c}, \label{mfp/vortex}
\end{equation}
i.e., $\ell \sim r_{{\rm L}}$
where $r_{{\rm L}}$ is the Larmor radius. In the limit $\omega _{c}\tau \gg
1 $, the vortex mean free path $\ell _{v}$ becomes shorter than the impurity
mean free path $\ell _{{\rm imp}}=v_{F}\tau $ so that the delocalized
excitations scatter on vortices more frequently than on impurities and thus
come to equilibrium with moving vortices. A similar consideration also
applies to localized excitations: In the limit $\omega _{0}\tau \gg 1$
interaction with a vortex is more effective than relaxation on impurities,
and the excitations come to equilibrium with the moving vortex.

In the intermediate case when $\omega _{c}\tau \rightarrow 0$ but $\omega
_{0}\tau \sim 1$, only the localized excitations are out of equilibrium. 
The excitations with energies above the energy gap $\Delta
_{\infty }$ have $\ell _v \gg \ell _{\rm imp}$ and are now in equilibrium 
with the heat bath: both $\gamma _{{\rm O}%
}^{\prime }=0$, and $\gamma _{{\rm H}}^{\prime }=0$. As a result these
excitations do not influence considerably the vortex motion. 
Delocalized excitations, however, do affect the vortex motion if $\omega
_{c}\tau $ is comparable with unity. Finally, in the limit $\omega _{0}\tau
,\omega _{c}\tau \ll 1$ the transverse force Eq. (\ref{force/tr}) disappears.
The longitudinal force dominates: the vortex dynamics is dissipative.

In general, we can identify several contribution to the transverse force.
One can present the full transverse force Eq. (\ref{force/tr}) in the form 
\begin{equation}
{\bf F}_{\perp} =-{\pi }N_s[ {\bf v}_{L}\times\hat{{\bf z}}]- {\pi }N_n[{\bf %
v}_{L}\times\hat{{\bf z}}]+{\bf F}_{{\rm sf}} .  \label{Fperp/presentation}
\end{equation}
The force 
\begin{eqnarray}
{\bf F}_{{\rm sf}}&=&{\pi }N \left\langle\!\!\!\left\langle 
\int {\frac{{d\mu }}{2}}\,\frac{\partial {\sl f}%
^{(0)}} {\partial \epsilon } \frac{\omega _0}{%
\omega _0^2\tau _0^2+1} \right\rangle\!\!\!\right\rangle [{\bf v}_{L}\times%
\hat{{\bf z}}]  \nonumber \\
&&+\pi N\left(1-\tanh\frac{\Delta _\infty}{2T}\right) \frac{1}{\omega
_c^2\tau ^2+1} [{\bf v}_{L}\times\hat{{\bf z}}]  \label{force/sf}
\end{eqnarray}
is called the spectral flow force%
\index{spectral flow force} \cite{KopVol/spflow}.

The spectral flow force ${\bf F}_{{\rm sf}}$ is due to the momentum flow
from the Fermi sea of normal excitations%
\index{excitations} to the moving vortex via the gapless spectral branch
going through the vortex core%
\index{vortex core} from negative to positive energies \cite{specflow},\cite
{specflow1}.  Due to
the time dependent angular momentum of the excitations $\mu (t)=[({\bf r}-%
{\bf v}_{L}t)\times {\bf p}]\cdot 
\hat{{\bf z}}$, there appears a flow (with the velocity $\partial \mu
/\partial t$) of spectral levels characterized by the angular momentum $\mu $%
. Each particle on a level carries a momentum $p_{F}$. The momentum transfer
to the vortex is effective if the quasiparticle relaxation occurs quickly:
the factor $(\omega _{0}^{2}\tau _{0}^{2}+1)^{-1}$ in the first line of Eq. (%
\ref{force/sf}) accounts for the relaxation on localized levels: the
relaxation and hence the momentum transfer is complete for $\omega _{0}\tau
\ll 1$, and vanishes in the opposite limit. The first term in Eq. (\ref
{force/sf}) thus describes a disorder--mediated momentum flow along the
anomalous chiral branch $E_{0}\left( \mu \right) $ for energies below the
gap. The second line in Eq. (\ref{force/sf}) accounts for the spectral flow
for energies above the gap. The corresponding factor which takes into
account the relaxation rate is $(\omega _{c}^{2}\tau ^{2}+1)^{-1}$.

Note that ${\bf F}_{\rm env}$ is calculated in the reference frame where
the normal component is at rest ${\bf v}_n=0$ for stationary vortices.
In superconductors, the normal velocity is always zero in equilibrium,
the excitations being at rest in the reference frame associated with
the crystal lattice. In superfluid $^{3}$He, the situation is similar: the
normal component has a rather large viscosity so that it is normally at
rest in the container frame of reference.
For a nonzero ${\bf v}_{n}$, we would need to replace ${\bf v}%
_{L}$ with ${\bf v}_{L}-{\bf v}_{n}$ everywhere in ${\bf F}_{\rm env}$.
Simultaneously, the Lorentz force would also include the contribution
due to the normal velocity.

Let us now turn to the force balance equation (\ref{forcebalance}). 
In presence of an electric field, the transport current is not entirely due 
to a supercurrent, a part of it being carried by delocalized quasiparticles.
Far from the vortex core, the quasiparticle current is
\begin{equation}
{\bf j}^{\left( {\rm qp}\right) } = -2\nu \left( 0\right) e
\int_{\epsilon >\Delta _\infty} {\bf v}_{\perp}g{\sl f}_{1}\,
d\epsilon \frac{d\Omega _{\bf p}}{4\pi } \label{j-qp}
\end{equation}
where $\nu (0)$ is the single--spin normal density of states, 
and $d\Omega _{\bf p}$ is 
the elementary solid angle in the direction of the momentum ${\bf p}$.
Using Eqs. (\ref{f1loc/sph}) and (\ref{gammas/del}) we find
\begin{equation}
{\bf j}^{\left( {\rm qp}\right) } = N_n e\gamma ^\prime _{\rm H}{\bf v}_L 
+N_n e\gamma ^\prime _{\rm O}\left[{\bf z}\times {\bf v}_L\right]
\label{j-qp1}
\end{equation}
where the density of normal quasiparticles is
\begin{equation}
N_n=N\int _{\epsilon >\Delta _\infty} g\frac{\partial {\sl f}^{(0)}}
{\partial \epsilon} \, d\epsilon .
\end{equation}

Writing the transport current as ${\bf j}_{{\rm tr}}=N_{s}e{\bf v}%
_{s}+{\bf j}^{\left( {\rm qp}\right) }$ we get the force balance Eq.
(\ref{forcebalance}) in the form 
\begin{equation}
{\bf F}_{{\rm M}}+{\bf F}_{\rm L}^{({\rm qp})}
+{\bf F}_{{\rm I}}+{\bf F}_{{\rm sf}}+{\bf F}_{\parallel }=0 .
\label{forcebal/Magnus}
\end{equation}
Here 
\[
{\bf F}_{{\rm M}}={\pi }N_{s}\left[ \left( {\bf v}_{s}-{\bf v}_{L}\right)
\times 
\hat{{\bf z}}\right] 
\]
is the Magnus force%
\index{Magnus force};
\begin{equation}
{\bf F}_{\rm L}^{({\rm qp})}=
\frac{{\Phi _{0}}}{c}[{\bf j}^{({\rm qp})}\times \hat{{\bf z}}]{\rm sign}%
\,(e)  \label{Lorentzforce/qp}
\end{equation}
is the Lorentz force from the quasiparticle current Eq. (\ref{j-qp1}), and
\[
{\bf F}_{{\rm I}}={\pi }N_{n}\left[ \left( {\bf v}_{n}-{\bf v}_{L}\right)
\times 
\hat{{\bf z}}\right] 
\]
is called the Iordanskii force \cite{Iordanskii}. The Iordanskii force
is the counterpart of the Magnus force for
normal excitations. We have included the normal velocity ${\bf v}_{n}$ to
make this similarity more transparent. 

The spectral flow force vanishes in the limit $\tau \rightarrow \infty $
when both $\omega _{0}\tau \gg 1$ and $\omega _{c}\tau \gg 1$. In this
limit the transverse force is given by Eq. (\ref{Ftransv/superclean}). 
The quasiparticle current is ${\bf j}^{({\rm qp})}=N_n e{\bf v}_L$ so that
the force from the quasiparticle current compensates the Iordanskii force.
The force balance Eq. (\ref{forcebal/Magnus}) reduces to
${\bf F}_{\rm M}=0$. The vortex thus moves with
the superfluid velocity ${\bf v}_L={\bf v}_s$. As follows from 
Eq. (\ref{distr/ideal}), quasiparticles
also have a velocity ${\bf v}_L$ so that all the particles move together
which amounts to the total current as in Eq. (\ref{jtr/superclean}).

On the contrary, the spectral flow force has its maximum value for moderately 
clean limit, $\omega _{0}\tau \ll 1$. Equation (\ref{force/sf}) gives in this
limit 
\[
{\bf F}_{{\rm sf}}={\pi }N[{\bf v}_{L}\times \hat{{\bf z}}]. 
\]
This completely compensates the first two terms in Eq. (\ref
{Fperp/presentation}), i.e, the Iordanskii force and the part of the Magnus
force that contains the vortex velocity; the transverse force vanishes.
The quasiparticle current vanishes even faster because 
$\omega _c\ll \omega _0$.
The Lorentz force is balanced only by a friction force ${\bf F}_{\parallel }$%
. As a result, the dissipative dynamics is restored.

\subsection{Low--field limit and  superfluid $^3$He}

The low--field limit when $\omega _c\tau \ll 1$ is most practical
for superconductors. Moreover, this regime is realized in electrically
neutral superfluids such as $^3$He. At the first glance, it is simply 
because $\omega _c$ vanishes together with the charge of carriers. However,
this is not completely correct. In fact, to estimate a deviation from 
equilibrium of delocalized excitations in this case one has again to compare
the mean free path of excitations with their mean free path with respect
to scattering by vortices. Keeping in mind that the vortex density is 
$n_v=2\Omega /\kappa$ where $\Omega$ is an angular velocity of a
rotating container and $\kappa =\pi /m$ is the circulation quantum,
Eq. (\ref{mfp/vortex}) gives $\ell_v\sim v_F/\Omega $.
We observe that the cyclotron frequency is replaced with the rotation velocity
in a full compliance with the Larmor theorem. The ratio of the 
particle--particle mean free path $\ell $ to the vortex mean free path is
$\ell /\ell _v \sim \Omega \tau$. With the practical rotation velocity
$\Omega$ of a few radians per second one always has
$\ell _v$ exceedingly larger than $\ell$. Delocalized
excitations are thus at rest in the container frame.

Consider this regime in more detail. Since delocalized excitations are  
in equilibrium the quasiparticle current vanishes. The force balance becomes
\begin{equation}
{\bf F}_{{\rm M}}
+{\bf F}_{{\rm I}}+{\bf F}_{{\rm sf}}+{\bf F}_{\parallel }=0
\label{forcebal/He3}
\end{equation}
where the spectral flow force is
\begin{eqnarray}
{\bf F}_{{\rm sf}}&=&{\pi }N \left\langle\!\!\!\left\langle 
\int {\frac{{d\mu }}{2}}\,\frac{\partial {\sl f}%
^{(0)}} {\partial \epsilon } \frac{\omega _0}{%
\omega _0^2\tau _0^2+1} \right\rangle\!\!\!\right\rangle [{\bf v}_{L}\times%
\hat{{\bf z}}]  \nonumber \\
&&+\pi N\left(1-\tanh\frac{\Delta _\infty}{2T}\right) 
[{\bf v}_{L}\times\hat{{\bf z}}] . \label{force/sf/3He}
\end{eqnarray}

It is interesting to observe that the spectral flow force from delocalized
states in this case is related to the anomalous contribution to the
transverse vortex cross section for scattering of delocalized
quasiparticles. The vortex cross sections%
\index{vortex cross section} were calculated in Refs.  \cite{KopKr2,GalSo}. 
The transverse cross section is 
\begin{equation}
\sigma _{\perp }=%
\frac{\pi }{p_{\perp }}\left[ \frac{\epsilon }{\sqrt{\epsilon ^{2}-\Delta
_{\infty }^{2}}}-1\right]  \label{cr-sec}
\end{equation}
per unit vortex length. Note that the transport cross section which is 
responsible for the scattering contribution
to the longitudinal force is vanishingly small in the semi--classical limit.
Inserting Eq. (\ref{cr-sec}) into the expression for
the force exerted on the vortex by scattered excitations%
\index{excitations}, 
\begin{equation}
{\bf F}_{\perp }^{({\rm sc})}=\int_{\epsilon >\Delta _{\infty}}%
\frac{d\epsilon }{2}\frac{\partial {\sl f}^{(0)}}{\partial \epsilon }\int 
\frac{dp_{z}}{2\pi }p_{\perp }^{3}\sigma _{\perp }[\hat{{\bf z}}\times {\bf v%
}_{L}],  \label{force/scat}
\end{equation}
we recover those contributions to Eq. (\ref{Fperp/presentation}) which are due
to the normal excitations, namely the term $-\pi N_n \left[{\bf v}_L\times 
\hat{\bf z}\right]$ and the last term in the spectral flow force Eq. 
(\ref{force/sf/3He}).

The first term in Eq. (\ref{cr-sec}) corresponds to the cross section of a
vortex in a Bose superfluid%
\index{Bose!superfluid} \cite{Sonin} 
\begin{equation}
\sigma _{\perp }^{\prime }=2\pi /m_{B}v_{g}  \label{cr-sec/Bose}
\end{equation}
where $v_{g}$ is the group velocity which, in our case, is defined by Eq. (%
\ref{v-group}), and $m_{B}$ is the mass of a Bosonic atom. Note that, in our
case, $m_{B}=2m$. The corresponding part of the transverse cross section can
be easily obtained from the semi--classical description. Indeed, the Doppler
energy due to the vortex velocity is ${\bf p}\cdot {\bf v}_{s}={\bf p}%
\cdot \nabla \chi /m_{B}$. Its contribution to the quasiparticle action is
\[
A=-%
\frac{1}{m_{B}}\int {\bf p}\cdot \nabla \chi \,dt=-\frac{p}{m_{B}v_{g}}\int 
\frac{\partial \chi }{\partial s}\,ds=-\frac{p}{m_{B}v_{g}}\delta \chi .
\]
Here $s$ is the coordinate along the particle trajectory as in
Fig. \ref{vort-coor}, and $\delta \chi $
is the variation of the order parameter phase along the trajectory. The
change in the transverse momentum of the particle is $\delta p_{\perp}=\partial
A/\partial b$ hence the transverse cross section becomes 
\[
\sigma _{\perp }^{\prime }=\int \frac{\delta p_{\perp}}{p}\,db
=\frac{A_{+}-A_{-}}{p} 
\]
where $A_{\pm }$ is the action along the trajectory passing on the left (right)
side of the vortex. 
Since $\delta \chi _{+}-\delta \chi _{-}=-2\pi $ we recover Eq. (%
\ref{cr-sec/Bose}). With this expression for the cross section, Eq. (\ref
{force/scat}) gives the Iordanskii force ${\bf F}_{{\rm I}}=-\pi N_{n}[{\bf v%
}_{L}\times \hat{{\bf z}}]$.

The second term in Eq. (\ref{cr-sec}) originates \cite{KopKr2} from the 
fact that here, as
distinct from the situation in a Bose superfluid, the phase of the
single-particle wave function changes by $\pi $ upon encircling the vortex,
while it is the order parameter phase which changes by $2\pi $. It is this
singularity produced by the vortex in the single-particle wave function
which results in the anomalous contribution to the cross section in Eq. (\ref
{cr-sec}). Inserted into Eq. (\ref{force/scat}), it exactly reproduces the
second term in Eq. (\ref{force/sf/3He}). We
see that the spectral flow force is related to a single--particle anomaly
associated with the vortex.

\section{Vortex momentum\label{sect/mass}}

The vortex mass%
\index{vortex mass} in superfluids and superconductors has been a long
standing problem in vortex physics and remains to be an issue of
controversies. There are different approaches to its definition. 
In early works on this subject, the vortex mass was determined through
an increase in the free energy of a superconductor calculated as an
expansion in slow time derivatives of the order parameter. The quasiparticle
distribution was assumed to be essentially as in equilibrium. First
used by Suhl \cite{Suhl} (see also \cite{Duan}) this approach yields the 
mass of the order of {\it one quasiparticle mass}%
\index{electron!bare mass} (electron, in case of superconductor) per atomic
layer. Another approach consists in calculating an
electromagnetic energy $E^2/8\pi$ which is proportional to the
square of the vortex velocity. This gives rise to the so
called electromagnetic mass \cite{Coffey} which, in good metals, is of the
same order of magnitude (see Ref. \cite{Blatter} for a review).

A crucial disadvantage of the above definitions of the vortex mass is that
they do not take into account the kinetics of excitations disturbed by a moving
vortex. We shall see that the inertia of excitations contributes much
more to the vortex mass than what the old calculations predict.
The kinetic equation approach described here is able to incorporate
this effect. To implement this method we find the force necessary to support
an unsteady vortex motion. Identifying then the contribution to the force
proportional to the vortex acceleration, one defines the vortex mass as a
coefficient of proportionality. This method was first applied for vortices
in superclean superconductors in Ref. \cite{Kop/mass} and then was 
used by other authors (see for example, Refs. \cite{KS,Simanek}). The 
resulting mass
is of the order of the {\it total mass of all electrons} within the area
occupied by the vortex core%
\index{vortex core}. We will refer to this mass as to the {\it dynamic mass}%
. Since the dynamic mass originates from the inertia of excitations 
localized in the vortex core%
\index{excitations!in the vortex core} it can also be calculated through the
{\it momentum carried by localized excitations} \cite{Vol/mass}.  
We shall see that dynamic mass displays a nontrivial
feature: it is a tensor whose components depend on the quasiparticle mean
free time%
\index{mean free time}. In $s$-wave superconductors,%
\index{superconductors!s-wave} this tensor is diagonal in the superclean
limit. The diagonal mass decreases rapidly as a function of the mean free
time, and the off-diagonal components dominate in the moderately clean
regime. These results were obtained in Ref. \cite{KopVin}. Our results agree
with the previous work \cite{Kop/mass,KS,Vol/mass} in the limit $\tau \to
\infty$.

\subsection{Equation of vortex dynamics}

To introduce the vortex momentum we consider a non--steady motion of a
vortex such that its acceleration is small. We again start with the
delocalized excitations. Multiplying Eq. (\ref{kin/local}) by ${\bf p}%
_{\perp }/2$ and summing up over all the quantum numbers, we obtain 
\begin{equation}
{\bf F}_{{\rm env}}^{\left( {\rm loc}\right) }={\bf F}_{{\rm coll}}^{\left( 
{\rm loc}\right) }-%
\frac{\partial {\bf P}^{\left( {\rm loc}\right) }}{\partial t}-\pi N\tanh
\left( \frac{\Delta _{\infty }}{2T}\right) [{\bf v}_{L}\times \hat{{\bf z}}]
\label{Fenv1}
\end{equation}
where the l.h.s. of Eq. (\ref{Fenv1}) is the force from the environment on a
moving vortex Eq. (\ref{force/boltz}).

The first term in the r.h.s. of Eq. (\ref{Fenv1}) is the force exerted on
the vortex by the heat bath via excitations localized in the vortex core: 
\begin{equation}
{\bf F}_{{\rm coll}}^{\left( {\rm loc}\right) }=-\frac{1}{2}\sum_{n}\int
{\bf p}_{\perp }\left( \frac{\partial f}{\partial t}\right) _{\rm coll} \,
\frac{dp_{z}}{2\pi }\,\frac{d\alpha \,d\mu }{2\pi }.
\end{equation}
The second term in the r.h.s. of (\ref{Fenv1}) is the change in the vortex
momentum%
\index{vortex momentum}  
\begin{equation}
{\bf P}^{\left( {\rm loc}\right) }=-%
\frac{1}{2}\sum_{n}\int {\bf p}_{\perp }f_{1}\,
\frac{dp_{z}}{2\pi }\, \frac{d\alpha \,d\mu }{2\pi } .  \label{momentum-loc}
\end{equation}

We turn now to delocalized excitations. Multiplying Eq. (\ref{kin/deloc}) by 
${\bf p}_{\perp }/2$ again and summing over all the states we find 
\begin{eqnarray}
{\bf F}_{{\rm env}}^{\left( {\rm del}\right) } &=&{\bf F}_{{\rm coll}%
}^{\left( {\rm del}\right) }-\frac{\partial {\bf P}^{\left( {\rm del}\right)
}}{\partial t}+\frac{1}{2}\sum_{n}\int \frac{dp_{z}}{2\pi }\frac{d\alpha }{%
2\pi }\frac{\partial {\sl f}^{\left( 0\right) }}{\partial \epsilon }\frac{%
\omega _{c}}{g}{\bf p}_{\perp }\left( {\bf p}_{\perp }\cdot \left[ {\bf \hat{%
z}}\times {\bf v}_{L}\right] \right)  \nonumber \\
&=&{\bf F}_{{\rm coll}}^{\left( {\rm del}\right) }-\frac{\partial {\bf P}%
^{\left( {\rm del}\right) }}{\partial t}-\pi N\left[ 1-\tanh \left( \frac{%
\Delta _{\infty }}{2T}\right) \right] [{\bf v}_{L}\times \hat{{\bf z}}] .
\label{Fenv3}
\end{eqnarray}
Here 
\[
{\bf F}_{{\rm coll}}^{\left( {\rm del}\right) }=-\frac{1}{2}\sum_{n}\int
{\bf p}_{\perp }\left( \frac{\partial f}{\partial t}\right) _{{\rm coll}}\,
\frac{dp_{z}}{2\pi }\,\frac{d\alpha}{2\pi}=\int_{\epsilon 
>\Delta _{\infty }}{\bf p}_{\perp }\frac{{\sl f}_{1}}{\omega _{c}\tau }
\, \frac{dp_{z}}{2\pi }\frac{d\alpha }{2\pi }\, d\epsilon  
\]
is the force from the heat bath, and the corresponding contribution to the
vortex momentum is 
\begin{equation}
{\bf P}^{\left( {\rm del}\right) }=-\frac{1}{2}\sum_{n}\int 
{\bf p}_{\perp }f_{1}\, \frac{dp_{z}}{2\pi }\, \frac{d\alpha}{2\pi }
=-\int_{\epsilon >\Delta _{\infty }}{\bf p}_{\perp }\frac{g}{\omega _{c}}f_{1}
\, \frac{dp_{z}}{2\pi }\frac{d\alpha}{2\pi }\, d\epsilon .
\label{momentum-del}
\end{equation}
The total momentum is ${\bf P}={\bf P}^{\left( {\rm loc}\right) }+{\bf P}%
^{\left( {\rm del}\right) }$.

After a little of algebra, the total force from the heat bath ${\bf F}_{{\rm %
coll}}={\bf F}_{{\rm coll}}^{\left( {\rm loc}\right) }+{\bf F}_{{\rm coll}%
}^{\left( {\rm del}\right) }$ can be written in the form 
\[
{\bf F}_{{\rm coll}}={\bf F}_{{\rm sf}}+{\bf F}_{\parallel } 
\]
where the friction%
\index{friction force} and spectral flow forces%
\index{spectral flow force} are determined by Eqs. (\ref{force/par}) and (%
\ref{force/sf}), respectively.
The total force from environment Eqs. (\ref{Fenv1}) and (\ref{Fenv3}) takes
the form 
\begin{equation}
{\bf F}_{{\rm env}}={\bf F}_{{\rm sf}}+{\bf F}_{\parallel }-%
\frac{\partial {\bf P}}{\partial t}-\pi N[{\bf v}_{L}\times \hat{{\bf z}}].
\label{Fenv2}
\end{equation}
This equation agrees with Eq. (\ref{Fperp/presentation}) for a steady
motion of vortices.

The equation of vortex dynamics is obtained by variation of the
superconducting free energy%
\index{free energy} plus the external field energy with respect to the
vortex displacement. The variation of the superfluid free energy gives the
force from the environment, ${\bf F}_{\rm env}$, while the variation of
the external field energy produces the external Lorentz force%
\index{Lorentz force}. In the absence of pinning the total energy is
translationally invariant. Therefore, the requirement of zero variation of
the free energy gives again the condition 
${\bf F}_{\rm L}+{\bf F}_{{\rm env}}=0$
in the form the force balance. Using our expression for ${\bf F}_{{\rm env}}$%
, the force balance can now be written in the form similar to Eq. (\ref
{forcebal/Magnus}) 
\begin{equation}
{\bf F}_{{\rm M}}+{\bf F}_{\rm L}^{({\rm qp})}+
{\bf F}_{{\rm I}}+{\bf F}_{{\rm sf}}+{\bf F}_{\parallel}= 
\frac{\partial {\bf P}}{\partial t}  \label{forcebal/momentum}
\end{equation}
where the r.h.s. contains the time derivative of the vortex momentum due to
a non--steady vortex motion. For a steady vortex motion, Eq. (\ref
{forcebal/momentum}) reduces to Eq. (\ref{forcebal/Magnus}). The physical
meaning of the Eq. (\ref{forcebal/momentum}) is simple. The l.h.s. of this
equation accounts for all the forces acting on a moving straight vortex
line. The r.h.s. of Eq. (\ref{forcebal/momentum}) comes
from the inertia of excitations and is
identified as the change in the vortex momentum. The definition of Eq. (\ref%
{momentum-loc}) is similar to that used in Refs. \cite{Stone,Vol/mass}. Note
that the delocalized quasiparticles do not contribute to the vortex momentum
in the limit $\omega _c\tau \ll 1$ because they are in equilibrium with the
heat bath.

\subsection{Vortex mass}

Having defined the vortex momentum, we calculate the vortex mass. The
distribution function%
\index{distribution function} is given by Eq. (\ref{f1loc/sph}). 
The vortex momentum becomes $P_{i}=M_{ik}u_{k}$; it has both
longitudinal and transverse components with respect to the vortex velocity.
For a vortex with the symmetry not less than the four-fold, the effective
mass tensor per unit length is $M_{ik}=M_{\parallel }\delta _{ik}-M_{\perp
}e_{ikj}{%
\hat{z}}_{j}$ where $M_{\parallel }=M_{\parallel \,e}+M_{\parallel \,h}$ and 
$M_{\perp }=M_{\perp \,e}-M_{\perp \,h}$. Each component contains
contributions from localized and delocalized states such that
$M_{\parallel ,\perp}=M_{\parallel ,\perp}^{({\rm loc})}+
M_{\parallel ,\perp}^{({\rm del})}$ where
\begin{equation}
M_{\parallel \,e,h}^{({\rm loc})}
=\frac{1}{4}\sum_{n}\int p_{\perp }^{2}\gamma _{\rm H}\,
\frac{df^{(0)}}{d\epsilon }\,
\frac{dp_{z}}{2\pi }\, \frac{d\mu \, d\alpha }{2\pi } 
= \pi N^{(e,h)}\left\langle\!\!\!\left\langle \sum_{n}
\int \gamma _{\rm H}\, \frac{df^{(0)}}{d\epsilon }\,
\frac{d\mu }{2 }\right\rangle\!\!\!\right\rangle 
\label{mass/tens/loc}
\end{equation}
and
\begin{equation}
M_{\parallel \,e,h}^{({\rm del})}=
\frac{1}{2}\int_{\epsilon >\Delta _\infty} p_{\perp }^{2}\gamma _{\rm H}%
^{\prime }\, \frac{g}{\omega _c}\,
\frac{df^{(0)}}{d\epsilon }\,\frac{dp_{z}}{2\pi }\, \frac{d\alpha }{2\pi }\,
d\epsilon  .  \label{mass/tens/del}
\end{equation}
The same expression holds for $M_{\perp \,e,h}$ where $\gamma _{{\rm H}}$ is
replaced with $\gamma _{{\rm O}}$. The indexes $e,h$  indicate the
corresponding momentum integrations over the electron and hole parts of the 
Fermi surface%
\index{Fermi!surface}, respectively. Only the branch with $n=0$ gives the
contribution to the transverse mass because $\gamma _{{\rm O}}$ is odd in $%
\mu $ for $n\ne 0$.

If the vortex
acceleration is slow, one can use expressions 
Eqs. (\ref{gammas}), (\ref{gammas/del})
for a steadily moving vortex to calculate the vortex inertia. 
Consider first the contribution of the states with $|\epsilon|>\Delta _\infty$.
Since $\gamma _{\rm H,O}^\prime$ do not depend on energy and momentum, 
Eq. (\ref{mass/tens/del}) gives 
\[
M_{\parallel, \perp}^{({\rm del})}
=\frac{\pi N_n^{(e,h)}}{\omega _c} \gamma _{\rm H,O}^\prime
=mN_n^{(e,h)}S_0 \gamma _{\rm H,O}^\prime
\]
The contribution of the delocalized states decreases as $\omega _c\tau$
gets smaller. In the limit of vanishing $\omega _c\tau$ the vortex mass
is determined by localized excitations. The localized excitations dominate
also at low temperatures $T\ll \Delta _\infty$. One has in this case
\[ 
M_{\parallel\, e,h}=\pi N^{(e,h)} 
\left\langle\!\!\!\left\langle \frac{\gamma _{\rm H}}
{\omega _0}\right\rangle\!\!\!\right\rangle,\;
M_{\perp\, e,h}=\pi N^{(e,h)} \left\langle\!\!\!\left\langle 
\frac{\gamma _{\rm O}}{\omega _0}\right\rangle\!\!\!\right\rangle .
\]

For $\omega _c\tau \gg 1$ the mass tensor for delocalized states is
diagonal $M_{ik}=M_\parallel \delta _{ik}$ where
$M_\parallel ^{({\rm del})}=mN_nS_0$; it is equal to the mass of normal 
particles in the area occupied by the vortex.
The mass tensor for localized states becomes diagonal in the superclean 
limit where $T_c^2\tau /E_F \gg 1$ with
$M_{\parallel}^{({\rm loc})} \sim 
\pi N\left\langle\!\left\langle \omega _0^{-1}\right\rangle\!\right\rangle
\sim \pi \xi _{0}^{2}mN$.
This is the mass of all electrons in the area 
occupied by the vortex core.
The mass decreases with $\tau $. In the moderately clean regime 
$T_c^2\tau /E_F \ll 1$ where 
$\omega _n\tau \ll 1$, the diagonal component vanishes as $\tau
^{2}$, and the mass tensor is dominated by the off-diagonal part.

We should emphasize an important point that, in contrast to a conventional
physical body, the mass of a vortex is not a constant quantity for a given
system: it may depend on the frequency $\omega $ of the external drive.
Indeed, for a nonzero $\omega $ we find from Eq. (\ref{kin/local}) 
\[
\gamma _{{\rm H}}=\frac{\omega _{n}^{2}\tau _{n}^{2}}{\omega _{n}^{2}\tau
_{n}^{2}+(1-i\omega \tau _{n})^{2}};~\gamma _{{\rm O}}=\frac{\omega _{n}\tau
_{n}(1-i\omega \tau _{n})}{\omega _{n}^{2}\tau _{n}^{2}+(1-i\omega \tau
_{n})^{2}}.
\]
As a result, all the dynamic characteristics of vortices including the
conductivity and the effective mass acquire a frequency dispersion. In the
limit $\tau \rightarrow \infty $, poles in $\gamma _{{\rm O,H}}$ appear at a
frequency equal to the energy spacing between the quasiparticle states in
the vortex core which gives rise to resonances in absorption of an external
electromagnetic field \cite{Kop/mass,Kop/dwres}.

\section{Conclusions}

All the dynamic characteristics of vortices such as
vortex friction, the flux flow conductivity, the Hall effect, and the vortex 
mass are determined by the mean free path of excitations which interact with
vortices. The key parameter is $\omega _0\tau$ where $\omega _0$ is the
interlevel spacing for the quasiparticle states in the vortex core.
For small $\omega _0\tau$, vortices experience viscous flow; the Lorentz
force is opposed by a friction force while the transverse force vanishes.
On the contrary, in a superclean regime when $\omega _0\tau \gg1$,
vortices move with a superflow as in an ideal fluid; the Hall angle is 
$\pi /2$, the friction force is zero, while the transverse force reaches 
its maximum value.
The vortex mass is a tensor. The longitudinal component dominates in the
superclean regime: it is the mass of all excitations in the vortex core.
On the contrary, the transverse component is the largest one in the
limit $\omega _0\tau\ll 1$.

I am pleased to acknowledge helpful discussions with E. Sonin
and G. Volovik. This work was supported by the Russian Foundation for Basic 
Research (Grant No. 99-02-16043).

\end{document}